\documentclass[aps, prl, %preprint, 
superscriptaddress, %tightenlines, 
nofootinbib, twocolumn]{revtex4}

\usepackage{amsmath}	% AMS Math Package
\usepackage{amsthm}	% Theorem Formatting
\usepackage{amssymb}	% Math symbols such as \mathbb
\usepackage{datetime}
\usepackage{graphicx}
\usepackage{color}
\usepackage{verbatim}

\definecolor{grey}{rgb}{0.4,0.4,0.4}
\definecolor{dullmagenta}{rgb}{0.4,0,0.4}
\definecolor{darkblue}{rgb}{0,0,0.4}
\definecolor{midblue}{rgb}{0,0,0.5}
\definecolor{midred}{rgb}{0.5,0,0}
\definecolor{orange}{rgb}{1,0.5,0}
\definecolor{lightbrown}{rgb}{0.75,0.5,0.25}
\definecolor{tan}{cmyk}{0.14,0.42,0.56,0}
\definecolor{djunglegreen}{cmyk}{0.99,0,0.52,0}
\definecolor{lightgreen}{rgb}{0,1,0}
\definecolor{olivegreen}{cmyk}{0.64,0,0.95,0.40}
\definecolor{midgreen}{rgb}{0.0,0.675,0.0}
\definecolor{darkgreen}{rgb}{0,0.5,0}

\newcommand{\q}{\quad}

\newcommand{\vs}{\vspace}

\renewcommand{\.}{\hspace{0.5mm}}

\newcommand{\Nrm}{\ensuremath{\mathrm{N}}}

\newcommand{\Prm}{\ensuremath{\mathrm{P}}}

\newcommand{\Xrm}{\ensuremath{\mathrm{X}}}

\newcommand{\crm}{\ensuremath{\mathrm{c}}}

\newcommand{\grm}{\ensuremath{\mathrm{g}}}

\newcommand{\mrm}{\ensuremath{\mathrm{m}}}

\newcommand{\srm}{\ensuremath{\mathrm{s}}}

\newcommand{\ie}{i.e.}

\newcommand{\cf}{c.f.}

\let\baraccent=\= % rename builtin command \= to \baraccent
\renewcommand{\=}[1]{\stackrel{#1}{=}} % for putting numbers above =

\theoremstyle{definition}

\theoremstyle{remark}

\smallskipamount
\settimeformat{ampmtime}

\usepackage{float}

\begin{document}

\title{Baryon number conservation in Bose-Einstein condensate black holes}

\author{Florian K{\"u}hnel}
\email{florian.kuhnel@fysik.su.se}
\affiliation{The Oskar Klein Centre for Cosmoparticle Physics,
	Department of Physics,
	Stockholm University,
	AlbaNova,
	106\.91 Stockholm,
	Sweden}

\author{Marit Sandstad}
\email{marit.sandstad@astro.uio.no}
\affiliation{Institute of Theoretical Astrophysics,
	University of Oslo,
	P.O.~Box 1029 Blindern,
	N-0315 Oslo,
	Norway}

\date{\formatdate{\day}{\month}{\year}, \currenttime}

\begin{abstract}
Primordial black holes are studied in the Bose-Einstein condensate description of space-time. The question of baryon-number conservation is investigated with emphasis on possible formation of bound states of the system's remaining captured baryons. This leads to distinct predictions for both the formation time, which for the naively natural assumptions is shown to lie between $10^{-12}\.\srm$ to $10^{12}\.\srm$ after Big Bang, as well as for the remnant's mass, yielding approximately $3 \cdot 10^{23}\.{\rm kg}$ in the same scheme. The consequences for astrophysically formed black holes are also considered.
\end{abstract}

\maketitle

%%%%%%%%%%%%%%%%%%%%%%%%%%%%%%%%%%%%%%%%%%%%%%%%%%%%%%%%%%%%%%
{\it Introduction\,---\,}
%\section{Introduction}
%\label{sec:Introduction}
%\setcounter{equation}{0}
%
%\noindent
The consideration of black holes as objects with important quantum-mechanical properties has been obvious since Hawking's discovery of their semi-classical evaporation \cite{Hawking:1974rv, Hawking:1974sw}. At approximately the same time, though, the issue of baryon number conservation in black holes was also considered \cite{Wheeler:1974, Carter:1974yx, Carr:1976zz}. However, this road has not been pursued much due to the assumption that baryon-number conservation should be broken or at least transcended \cite{Wheeler:1974} by the black hole. This and other related manifestations of the no-hair theorem have not been understood in any semi-classical approach.

A fully quantum proposal has been made by Dvali and Gomez to describe black holes, and other space-time geometries, as the results of certain peculiar configurations of a background Bose-Einstein condensate of gravitons \cite{DvaliGomez-N-Portrait, DvaliGomez, Dvali:2012rt} (see \cite{Flassig:2012re, Casadio:2015xva, Casadio:2015bna, Kuhnel:2014oja, Kuhnel:2015yka, mueckPT, Hofmann, Binetruy:2012kx, Kuhnel:2014gja, Casadio:2014vja, Brustein, Foit:2015wqa} for recent developments). Therein it is indeed possible to resolve all semi-classical paradoxes, in particular the one mentioned before. In this approach, any other species, like a baryon, which is captured by the black hole is strongly bound by the self-sustained bound state of condensed gravitons which make up the black hole. Now, the very mechanism which is responsible for Hawking radiation, namely quantum depletion, is also responsible for emission of any captured quantum, which is fully released over the life time of the black hole.

It is the aim of this paper to review the issue of baryon number conservation as well as the formation of related bound states in this novel corpuscular formulation of black holes. This shall be done in the context of black holes created in the very early Universe, \ie~primordial black holes \cite{Zel'dovich-Novikov1967, Hawking:1971ei}, but we will also consider briefly the consequences for black holes formed from the astrophysical collapses of very massive stars. Specifically we wish to consider possible formation of bound states of remaining baryons as first hypothesised in \cite{Carter:1974yx} thereby quantifying the consequences of the baryon conservation in the Bose-Einstein condensate considered in \cite{Dvali:2012rt}.

%%%%%%%%%%%%%%%%%%%%%%%%%%%%%%%%%%%%%%%%%%%%%%%%%%%%%%%%%%%%%%
{\it Primordial black holes\,---\,}
%\section{Primordial black holes}
%\label{sec:PBH}
%\setcounter{equation}{0}
%
%\noindent
Primordial black holes are black holes formed in the very early Universe \cite{Zel'dovich-Novikov1967, Hawking:1971ei}. They are generally assumed to form when a critical mass over-density crosses the horizon and can subsequently create a horizon-size black hole. These over-densities can theoretically stem from the extreme ends of the initial inflationary spectrum or can also be sourced by exotic early-Universe phenomena such as cosmic string loops or bubble collisions (\cf~\cite{Green:2014faa} for a recent review).

As these collapse more or less immediately after crossing the horizon, the initial mass $M_{*}$ of the black holes they form are given as the mass of a black hole with Schwarzschild radius equal to the horizon size at their formation time $t_{*}$. This can be found to be roughly of the order \cite{Carr:2005bd}:
\vs{-2mm}
\begin{align}
	M_{*}
		&\approx
								\frac{c^{3}\.t_{*}}{G_{\Nrm}}
		\approx
								10^{12}
								\left(
									\frac{ t_{*} }{ 10^{-23}\.\srm }
								\right)
								{\rm kg}
								\; ,
								\label{eq:PBHMass}
\end{align}
with $c$ being the speed of light, and $G_{\Nrm}$ is Newton's constant. It is assumed that primordial black holes will not accrete substantially, so that their evolution is determined more or less only by the Hawking evaporation process that they undergo. Thus from the evaporation of the black hole we can define a lifetime $t_{\rm life}$ given by \cite{Hawking:1974rv}:
\begin{align}
	t_{\rm life}
		%\approx
		%						\frac{\hslash\.c^{4}}{G_{\Nrm}^{2} M^{3}}
		\approx
								10^{71}
								\left(
									\frac{M_{*}}{M_{\odot}}
								\right)^{\!3}
								\srm
								\; ,
								\label{eq:ClassLifeTime}
\end{align} 
where $M_{\odot}$ is the solar mass.\footnote{For the extremely low mass black holes, the lifetime is shortened by the possibility of evaporation through additional particle species as the small black holes have higher temperature. For the black holes near the Planck mass $M \approx 10^{-5}\.\grm$, effects of the uncertainty principle also comes into play, and the lifetime is only given as an expectation value of the lifetime.} Since primordial black holes formed later then $t \approx 10^{-23}\.\srm$, \ie~of a mass $M > 10^{13}\.{\rm kg}$ have lifetimes longer than the current age of the Universe, primordial black holes are potential candidates for dark matter in the Universe. For most masses the possible fraction of the dark matter that can be in the form of primordial black holes are constrained by the non-detection of their expected observable effects (see \cite{Green:2014faa, Carr:2009jm, Capela:2013yf} for recent reviews).

%%%%%%%%%%%%%%%%%%%%%%%%%%%%%%%%%%%%%%%%%%%%%%%%%%%%%%%%%%%%%%
{\it Bose-Einstein Condensates\,---\,}
%\section{Bose-Einstein Condensates}
%\label{sec:BECIntro}
%\setcounter{equation}{0}
%
%\noindent
It has been suggested by Dvali and Gomez \cite{DvaliGomez-N-Portrait, DvaliGomez} that classical geometries, like black holes and de Sitter Universes, can be only the classical limit of quantum critical states of graviton Bose Einstein condensates. Here the defining length scale $L$ is given by the typical wavelength of the constituent gravitons and the number $N$ of gravitons in the coherent state is also defined by this length scale according to
\begin{align}
	N
		&\simeq
								\left(
									\frac{ L }{ L_{\Prm} }
								\right)^{\!2}
								\; ,
								\label{eq:N}
\end{align}
where $L_{\Prm}$ is the Planck length.
In the case of a black hole, the defining length scale is the Schwarzschild radius $R_{H} = 2\.G_{\Nrm}\.M / c^{2}$ of the resulting classical object, and hence the number of gravitons is given as
%\vs{-2mm}
\begin{align}
	N
		&\simeq
								\left(
									\frac{2\.G_{\Nrm}}{c^{2} L_{\Prm}}\.M
								\right)^{\!2}
								\; .
								\label{eq:NofM}
\end{align}
Generically, the number of gravitons in the condensate is very large, so the main process to perturb gravitons from their condensate state is given by graviton-to-graviton scattering.

Since the wavelength is also given by the characteristic length scale, the interaction strength of this interaction is $\alpha = 1 / N$. This means that the condensate is at a self-similar critical point, \ie~it remains at criticality even as gravitons are added or subtracted from the condensate. The graviton-to-graviton scattering, however, leads to a depletion of the soft ground-state gravitons of characteristic energy $\epsilon = 1 / \sqrt{N\,}$ according to
\begin{subequations}
\begin{align}
	\dot{N}
		&\simeq
								-\,
								N ( N - 1 )\.
								\alpha^{2}\.
								\frac{ \epsilon }{ \hslash }
		\simeq
								-\,
								\frac{c}{L_{\Prm}\.\sqrt{N\,}}
								\; ,
								\label{eq:dot-N(t)}
\end{align}
where the first factor $N ( N - 1 )$ is a combinatorial one, which is cancelled by $\alpha^{2} = 1 / N^{2}$ (coming from the two relevant vertices). Here and in the following we shall assume $N \gg 1$, and in turn ignore higher-order terms in $1 / N$. If the condensate in addition contains a number $N_{\Xrm} \ll N$ of X-particles, possibly of a conserved X-charge, these particles will also deplete from the condensate via scattering off gravitons, and their depletion will be given by (\cf~Eq.~(10) in Ref.~\cite{Dvali:2012rt})
%\vs{-2mm}
\begin{align}
	\dot{N}_{\Xrm}
		&\simeq
								-\,
								N N_{\Xrm}\.
								\alpha^{2}\.
								\frac{ \epsilon }{ \hslash }
		\simeq
								-\,
								\frac{c}{L_{\Prm}\.\sqrt{N\,}}\frac{N_{\Xrm}}{N}
								\; .
								\label{eq:dot-NX(t)}
\end{align}
\end{subequations}
Note that for $N_{\Xrm} \ll N$ the depletion rate \eqref{eq:dot-NX(t)} will be much smaller than that for the gravitons, given by \eqref{eq:dot-N(t)}. In the following we will take this conserved particle species to be baryonic, however the general formulae for the number and densities of the particles will be valid also for other species of conserved charge.

%%%%%%%%%%%%%%%%%%%%%%%%%%%%%%%%%%%%%%%%%%%%%%%%%%%%%%%%%%%%%%
{\it Depletion and formation of new bound state\,---\,}
%\section{Depletion and formation of new bound state}
%\label{sec:Main-Part}
%\setcounter{equation}{0}
%
%\noindent
Let us consider a system with $N_{B}$ baryons surrounded by $N$ gravitons in a quantum critical condensate state. We assume that initially $N \gg N_{B}$, and work to leading order in $1 / N$. 
The equation set spanned by Eqs.~(\ref{eq:dot-N(t)},b) is analytically soluble and the solutions read (\cf~Eqs.~(13) in Ref.~\cite{Dvali:2012rt})
\begin{subequations}
\begin{align}
	N( t )
		&\simeq
								N_{*}
								\left( 
									1
									-
									\frac{ 3 }{ 2 }
									\frac{ t - t_{*} }{ N_{*}^{3 / 2} }
								 \right)^{\!2 / 3}
								\; ,
								\label{eq:N(t)}
								\\[2mm]
	N_{B}( t )
		&\simeq
								N_{B*}
								\left( 
									1
									-
									\frac{ 3 }{ 2 }
									\frac{ t - t_{*} }{ N_{*}^{3 / 2} }
								 \right)^{\!2 / 3}
								\; ,
								\label{eq:NB(t)}
\end{align}
\end{subequations}
with $N_{*} = N( t_{*} )$ and $N_{B*} = N_{B}( t_{*} )$ being the number of gravitons and baryons at the formation time $t_{*}$. Approximately, total lifetime $t_{\rm life}$ can be found to be $t_{\rm life} \approx 2 / 3\,N_{*}^{3 / 2}$, which corresponds exactly to the classical lifetime given by Eq.~(\ref{eq:ClassLifeTime}). This lifetime is at first glance the same for both the graviton condensate and the baryons inside it.

Though quantum depletion causes the baryon number inside the condensate to decrease over time, the volume which the condensate spans decreases much faster, causing the density of the baryons to increases. Since the baryons can interact strongly through the strong force when they are close together, it is not unnatural to assume that when the baryons in the condensate reach the QCD-scale density, their mutual strong interactions will become comparable to their interactions with the gravitons, causing a phase transition in the baryon-graviton ensemble. More precisely, at a number density of approximately $n_{\crm} \equiv 1\.{\rm baryon} / {\rm fm}^{3}$ for the baryons we postulate that the state of the condensate changes into a stable state, no longer depleting. We dub the time when this possible phase transition occurs $t_{\crm}$.\footnote{Whether this exactly constitutes a stable state for the gravitons and not only the baryons, is very unclear as this is a physical situation very far from known theory or experimental reach. However, the definition of this time, or a nearby time as a transitioning time for the ensemble is natural.} This happens when
\begin{align}
	\frac{ N_{B}( t_{\crm} ) }{ \frac{ 4 \pi }{ 3 } R_{S}( t_{\crm} )^{3} }
		&=
								n_{\crm}
								\; ,
								\label{eq:nc}
\end{align}
yielding
\begin{align}
	t_{\crm}
		&\simeq
								t_{*}
								+
								\frac{2}{3}\.N_{*}^{3/2}
								\left(
									1
									-
									\frac{27\.N_{B*}^{3}}
									{64\.\pi^3 N_{*}^{9/2} n_{\crm}^3}
								\right)
								.
								\label{eq:tc}
\end{align}
This time is given only in terms of the initial numbers of gravitons and baryons, \ie~they are set at the time of formation of the black hole. We note that for sufficiently large initial values of baryon to graviton ratios, this time will become negative. We interpret this to mean that states that would otherwise form black holes for which the initial baryon number is too high, a black hole state will not form, and the state will be dominated by the strong force instead. That is, the second factor in the brackets of Eq.~\eqref{eq:tc} should not exceed one in order to ensure $t_{\crm} > 0$, \ie
\vs{-2mm}
\begin{subequations}
\begin{align}
	\frac{27\.N_{B*}^{3}}
	{64\.\pi^3 N_{*}^{9/2} n_{\crm}^3}
		&\leq
								1
								\; ,
								\label{eq:aa}
\end{align}
and thus possible black hole formation.

Also, as an absolute bound, at $t = t_{\crm}$ there should be at least one baryon left in the condensate for the new state reached to have any sensible meaning, \ie
\begin{align}
	\frac{9\.N_{B*}^{3}}
	{16\.\pi^{2}N_{*}^{3}\.n_{\crm}^{2}}
		&\geq
								1
								\; .
								\label{eq:bb}
\end{align}
\end{subequations}
Of course, in practice the latter bound will be much stricter. Below we investigate the consequences of these bounds and possible formation of a new stable state in the case of primordial black holes, and black holes formed from more astrophysical scenarios.

%\subsection{Primordial black holes}
%{\it Primordial black holes\,---\,}
Both \eqref{eq:aa} as well as \eqref{eq:bb} set limits on the earliest as well as on the latest time of formation of the primordial condensate. In order to investigate this, let us assume radiation domination during the relevant time interval. Then the comoving Hubble radius is given by $R_{H}( t ) = 2\.c\,\sqrt{t_{0}\,} \sqrt{t\,}$, where $t_{0}$ equals the age of the Universe. For getting a rough estimate of the total baryon number contained in a Hubble patch at the time $t_{*}$, we can utilize today's value of the baryon number density $n_{B, 0} \approx 0.2 / \mrm^{3}$, and estimate $N_{B*}$ via
\begin{subequations}
\begin{align}
	N_{B*}
		&\simeq
								n_{B, 0}\,\frac{ 4\.\pi }{ 3 } R_{H}( t_{*} )^{3}
								\; .
								\label{eq:NB*-estimate}
\end{align}
A horizon-size primordial BEC at $t = t_{*}$ approximately has mass as given by Eq.~\eqref{eq:PBHMass} and contains
\begin{align}
	N_{*}
		&\simeq
								4\.\left(\frac{M( t_{*} )}{M_{P}}\right)^{2}
		\approx
								10^{86}\.t_{*}^2[ \srm ]
								\label{eq:N*-estimate}
\end{align}
\end{subequations}
gravitons. Using Eqs.~(\ref{eq:NB*-estimate},b), we find from Eqs.~(\ref{eq:aa},b) that the time of condensate formation has to be within the interval
\begin{align}
	10^{-12}\.\srm
	\lesssim
	t_{*}
	\lesssim
	10^{12}\.\srm
								\label{eq:t*-interval}
\end{align}
to lead to a possible non-trivial end state where gravity is balanced by the strong force. Let us stress again that in the Bose-Einstein condensate framework baryon number is conserved in any (would-be) black hole state. Hence, because at times earlier than $ 10^{-12}\.\srm$ the baryon density is so high that the strong force counteracts the gravitational one, no black-hole-like bound state will form. This would affect the spectrum of any model of primordial black hole formation by removing the majority of the  low mass black holes from it.

Since the time of formation of the stable state increases very rapidly with lower initial baryon density essentially all primordial condensates which are produced during this interval constitute of the ones that lay on the boundary and form the new hypothesised bound state more or less immediately. These have
\vs{-2mm}
\begin{align}
	N_{B}
		&\simeq
								2 \cdot 10^{35}
	\q
	\text{and}
	\q
	N
		\simeq
								6 \cdot 10^{62}
								\label{eq:BEC-PBH-numbers}
\end{align}
baryons and gravitons, respectively, leading to a mass of
\begin{align}
	M
		&\simeq
								3 \cdot 10^{23}
								\.{\rm kg}
								\; .
								\label{eq:BEC-PBH-mass}
\end{align}
This is roughly an order of magnitude below the mass of the Earth. Since we do not expect primordial black holes or the new stable state condensate to form prior to the formation of these particular ones, we will expect a boost in formation of objects of exactly this size. Thus this framework provides a definite prediction of the remnant's mass, which can then be compared to the non-detection limits for primordial black holes of this size. The tightest bounds in this mass regime stem mainly from galactic microlensing studies such as the EROS \cite{Tisserand:2006zx} and MACHO \cite{Alcock:2001} surveys, or recent inspection of Kepler data \cite{Griest:2013aaa}.

%\subsection{Astrophysically formed black holes}
{\it Astrophysically formed black holes\,---\,}
For a black hole formed as an end state of a very massive star, the limits given by Eqs.~(\ref{eq:aa},b) will instead of setting limits on the formation time, rather set a limit on initial densities and mass of the star that collapses. Because the main part of the mass contained in the late stages of a very massive star is composed of baryons, we can find the initial number of baryons as:
\begin{align}
	N_{B*}
		&\approx
								\frac{ M }{ M_{p} }
		\approx
								10^{57}\.
								\frac{ M }{ M_{\odot} }
								\; ,
\end{align}
where $M_{p}$ is the proton mass, while the number of gravitons is still set by the mass of the star according to Eq.~\eqref{eq:NofM}. As the size of the black hole, and hence its Schwarzschild radius is given by the same initial mass of the late-stage star, we can simply compute the number of baryons per femtometer as a function of the mass. If it exceeds 1, the star will then presumably not collapse to a black hole, but instead form something else, possibly something like a quark star \cite{Itoh:1970uw}. 
The limit for this becomes:
\begin{align}
	M
		&\geq
								\frac{c^3}{4}
								\sqrt{\frac{3 \cdot 10^{57}}{2\pi\.M_{\odot}\.n_{\crm}}\,}\;
								G_{\Nrm}^{-3 / 2}
		\approx
								3\.M_{\odot}
								\; ,
\end{align}
which is roughly twice the normal Chandrasekhar limit \cite{Chandrasekhar:1931ih}, which is the normal bound on black hole formation for white dwarfs, but lies around the Tolman-Oppenheimer-Volkoff limit \cite{TOVlimit} for black hole formation from neutron stars. Thus the lightest candidates for black hole formation, may form other exotic states instead such as quark stars. However, since the time it takes for the QCD condensate bound state to form, Eq.~\eqref{eq:tc}, is very sensitive to slight changes in the mass. Only the stars that are very close to the bound might form this exotic state. Here, unless there is some mechanism to strip the lighter remnants of baryons, there are no mechanisms for overproduction of states on the bound, thus these will be very rare.

%%%%%%%%%%%%%%%%%%%%%%%%%%%%%%%%%%%%%%%%%%%%%%%%%%%%%%%%%%%
{\it Summary and Outlook\,---\,}
%\section{Summary and Outlook}
%\label{sec:Summary-and-Outlook}
%\setcounter{equation}{0}
%
%\noindent
In this work we have investigated the issue of baryon number conservation as well as the formation of related bound states in the corpuscular picture of black holes as a concretisation of the work done in \cite{Dvali:2012rt}. Here this has been done in the context of the black holes created in the very early Universe, \ie~primordial black holes, but we also considered briefly the consequences for the ones formed from astrophysical collapses of very massive stars.

Regarding the former, we found that the corpuscular framework leads to distinct predictions for both the formation time, which is shown to happen only after $10^{-12}\.\srm$ after Big Bang. At times earlier than this the baryon density, which is preserved in the graviton condensate framework, is so high that the strong force dwarfs gravity and no black-hole-like bound state will form. For black holes formed before $10^{12}\.\srm$ after the Big Bang the endstate we hypothesise is not a totally evaporated state, but rather a bound state of baryons and gravitons, where gravity is counteracted by the strong force. However, only the black holes formed very close to the $10^{-12}\.\srm$ will have reached the bound state by today; \ie~we expect the spectrum of primordial black holes of mass smaller than $~3 \times 10^{23}\.{\rm kg}$ to be essentially zero and then an enhancement of the spectrum for masses just above this bound with respect to predictions given in the standard paradigm. This means that even inflationary theories which predict peaks in observationally disallowed low mass primordial black holes might be viable within this framework. These statements are rather general, and apply to basically any model of primordial black-hole formation. Also the possible enhancement of the spectrum occurs at values of primordial black hole mass for which there are only weak observational constraints \cite{Carr:2009jm, Capela:2013yf}.

% Each given spectrum of over-densities entering the horizon is essentially cut below the minimum time at which gravity and the strong force balance. This implies the absence of any possibly dangerous overproduction, present in some inflation models, and thus increases the range of viable models of primordial black-hole production. 

Concerning astrophysical black holes, we have derived a bound on the minimal mass of such objects and found that it lies just above the Chandrasekhar limit.

%As our work is the first investigating the question of primordial black holes on the corpuscular picture of space-time

Many more things need to be investigated{\,---\,}in particular, the dynamics and structure of the graviton-baryon bound state, the exact nature of the phase transition\./\.cross-over when exceeding the critical baryon density, and the determination of the latter. Also, the formation mechanisms of primordial black holes should be re-investigated in the corpuscular description of space-time with emphasis on the effect of possible quantum corrections to the mass spectrum. More precise investigation of this will most likely result in shifts in the predictions for the remnant masses and bounds, however the general outline given here should still hold. Furthermore, keeping in mind that all of those objects contribute to the dark matter density, one should study also more exotic states which involve a much larger baryon-to-graviton ratio than the ones investigated in this work.\\% and a unabridged investigation of those objects might unveil more insights into it.

%%%%%%%%%%%%%%%%%%%%%%%%%%%%%%%%%%%%%%%%%%%%%%%%%%%%%%%%%%%
\acknowledgments
%{\it Acknowledgements\,---\,}
F.~K.~acknowledges supported from the Swedish Research Council (VR) through the Oskar Klein Centre, and thanks the Institute of Theoretical Astrophysics at the University of Oslo where part of this work as been performed. We thank Gia Dvali for valuable comments.

%%%%%%%%%%%%%%%%%%%%%%%%%%%%%%%%%%%%%%%%%%%%%%%%%%%%%%%%%%%%

\end{document}